\documentclass{aa}
\usepackage[obeyspaces]{url}
\usepackage{hangcaption} 
\usepackage{natbib} 
\usepackage{graphicx}
\usepackage{amssymb}

\begin{document}
\title{Pulsar science with the Five hundred metre Aperture Spherical Telescope} 
\author{R.~Smits \inst{1} \and D.R.~Lorimer \inst{2,3} \and M.~Kramer \inst{1,4} \and 
  R.~Manchester\inst{5} \and B.~Stappers\inst{1}
\and C.J.~Jin \inst{6} \and R.D.~Nan \inst{6} \and D.~Li \inst{7} }
\offprints{Roy Smits\\ \email{Roy.Smits@\allowbreak
  manchester.\allowbreak ac.uk}} 
\institute{
  University of Manchester, Jodrell Bank Centre for Astrophysics, Alan-Turing Building, Oxford Road, Manchester, M13 9PL, UK
  \and Department of Physics, Hodges Hall, West Virginia University,  Morgantown, WV 26506, USA 
  \and National Radio Astronomy Observatory, Green Bank Observatory, PO Box 2, Green Bank, WV 24944, USA
  \and Max-Planck-Institut f\"ur Radioastronomie, Bonn, Germany, Auf dem Huegel 69, 53121 Bonn, Germany
  \and Australia Telescope National Facility, CSIRO, PO Box 76, Epping NSW 1710, Australia
  \and National Astronomical Observatories, Chinese Academy of Sciences
  Chaoyang District, Datun Road, A.20, Beijing 100012, China
  \and Jet Propulsion Laboratory, California Institute of Technology, 4800
  Oak Grove Dr. Pasadena, CA 91109 }

\date{Received / Accepted } \abstract { With a collecting area of
  70\,000\,m$^2$, the Five hundred metre Aperture Spherical Telescope
  (FAST) will allow for great advances in pulsar astronomy. We have
  performed simulations to estimate the number of previously
    unknown pulsars FAST will find with its 19-beam or possibly
  100-beam receivers for different survey strategies. With the 19-beam
  receiver, a total of 5\,200 previously unknown pulsars could
  be discovered in the Galactic plane, including about 460 millisecond
  pulsars (MSPs). Such a survey would take just over 200 days
  with eight hours survey time per day. We also estimate that,
    with about 80 six-hour days, a survey of M31 and M33 could yield
  50--100 extra-Galactic pulsars. A 19-beam receiver would produce
  just under 500~MB of data per second and requires about 9
  tera-ops to perform the major part of a real time analysis. We also
  simulate the logistics of high-precision timing of MSPs with
  FAST. Timing of the 50 brightest MSPs to a signal-to-noise of
    500 would take about 24 hours per epoch.  }
\authorrunning{Smits et al.}  \titlerunning{Draft: Pulsar science with
  FAST} \maketitle

\section{Introduction}
FAST, the Five-hundred-metre Aperture Spherical Telescope, is an
Arecibo-style telescope currently under construction in
  China. FAST will be located in a karst depression in Guizhou
province at a latitude of about $+26\degr$ and is planned to be
completed around 2014 \citep{Nanb06, Nan08}.  With an illuminated
aperture of 300\,m diameter, it will be an order of magnitude more
sensitive than the steerable 100-metre telescopes at Effelsberg and
Green Bank and about twice as sensitive as Arecibo. Moreover, FAST
will cover a frequency range of 70\,MHz to 3\,GHz and it will be able
to observe at a zenith angle of 40 degrees without a significant loss
in gain \citep{jin08}. By illuminating only part of the 300-metre
aperture the zenith angle can be extended even further; at a zenith
angle of 60 degrees about half the aperture is still illuminated. With
its sensitivity and its ability to see the majority of the sky,
including a large part of the Galactic plane, FAST has excellent
potential for finding radio pulsars \citep[see also][]{nan06}.

Radio pulsars provide insights into a rich variety of physics and
astrophysics. Applications to date \citep[see e.g.][]{lk05} include
the study of the Milky Way, globular clusters, the evolution and
collapse of massive stars, the formation and evolution of binary
systems, the properties of super-dense matter, extreme plasma physics,
tests of theories of gravity and as cosmological probes. A goal for
the coming decade is the direct detection of low-frequency
gravitational waves via high-precision timing of an array of
millisecond pulsars (MSPs) \citep{Jenet05}. Such an array would be
greatly facilitated by further discoveries and timing of MSPs with
FAST.

The aim of this paper is to investigate the possibilities of FAST
for finding and timing radio pulsars. In \S 2 we present a
simulation of a FAST survey of the pulsar sky, where we look into the
survey speed and the number of pulsars that can be detected for
different receivers.  In \S 3 we present calculations to estimate the
data rates and computational requirements for a pulsar survey with
FAST. In \S 4 we look into the possibility for FAST to participate in
a global pulsar timing array effort. \S 5 contains a discussion of the
results, as well as a comparison between FAST and the Square Kilometre Array.

\section{Survey simulation}
In the past, the southern hemisphere has been the subject of extensive
pulsar surveys \citep{mld+96,lml+98,mlc+01}. FAST will be able to
complement these surveys in the northern hemisphere with an even
greater sensitivity. Initially, FAST will have a 19-beam receiver with
a frequency range of 1.1--1.5\,GHz and a system temperature of 20\,K
(not including the sky temperature). A possible future phased
array feed (PAF) is planned to have over 100 beams \citep{jin08} 
  with the same frequency range and a system temperature of 30\,K.
Table~\ref{tab:parameters} shows the parameters involved in the survey
simulation. The number of beams will determine the speed of the survey
and the dwell time per pointing. We therefore perform the simulations
as a function of both frequency and field of view (FoV).

\begin{table}
  \caption{Expected system parameters of FAST and the pulsar survey.}
  \begin{tabular}{ll}
    \hline
    \hline
    Parameter                     & Value \\
    \hline
    Gain                          & 16.5\,KJy$^{-1}$ \\
    System temperature 19-beam receiver   & 20\,K \\
    System temperature 100-beam receiver  &30\,K \\
    Centre frequency              & 1315\,MHz \\
    Bandwidth                     & 400\,MHz \\
    Number of frequency channels  & 9\,500 \\
    No. of polarisations          & 2 \\
    Beam FWHM                     & 3$^\prime$.4 \\
    No. of beams                  & 19 or 100 \\
    \hline
  \end{tabular}
  \label{tab:parameters}
\end{table}

\subsection{Simulation method}
We performed Monte Carlo simulations following \citet{Lorimer2006}
using the psrpop\footnote{\url{http://psrpop.sourceforge.net}.}
package \citep[cf.][]{Smits09}.  In their study, \citet{Lorimer2006}
used the results from recent surveys with the Parkes Multibeam system
to derive an underlying population of pulsars with an optimal set of
probability density functions for pulsar period ($P$), 1400-MHz radio
luminosity ($L$), Galactocentric radius ($R$) and height above the
Galactic plane ($z$). We make use of these results in our simulations
described below which use as a starting point model C$^\prime$ from
\citet{Lorimer2006}.

Our simulation procedure begins by generating a population of normal
pulsars which beam towards the Earth. Each pulsar is assigned a value
of $P$, $L$, $R$ and $z$ based on the assumed probability density
functions.  For the distributions in $P$, $R$ and $z$, we use the
distributions from C$^\prime$ of \citet{Lorimer2006}. To compute $L$,
we adopt the log-normal distribution found
by~\citet{Faucher2006}. This has the advantage of not requiring a
specific lower bound in $L$, as is the case for the power-law
luminosity models considered by \citet{Lorimer2006}.  

We compute intrinsic pulse widths using the following self-consistent
approach.  Assuming a simple geometry with circular beams of radius
$\rho$, the pulse width can be found from the inclination angle
between the pulsar spin and magnetic axis, $\alpha$, and the impact
parameter between the magnetic axis and the line of sight, $\beta$.
Following Kramer et al.~(1998), we use the empirical relationship
between $\rho$ and spin period $P$ (s) defined as follows:
\begin{displaymath}
\rho_{\rm model} = \left\{ 
    \begin{array}{ll}
      5.4^{\circ} P^{-1/2}
       & \mbox{if $P>30$~ms} \\
      31.2^{\circ} & \mbox{otherwise},
    \end{array} 
    \right.
\end{displaymath}
where the angle of $31.2^{\circ}$ is the value of the above expression
evaluated for $P=30$~ms. To model the scatter about this relationship
(see, e.g., Fig.~3.5 from Lorimer \& Kramer 2005), we use a Monte
Carlo approach to find a dithered beam radius
\begin{equation}
  \rho = 10^{\log_{10} (\rho_{\rm model}) + p},
\end{equation}
where $p$ is a random number drawn from a flat distribution 
in the range $-0.15 \leq p \leq 0.15$. This, somewhat arbitrary,
scheme dithers the beam radius logarithmically by a third of a decade
and provides a good match to the observed scatter.
The pulsar is assumed to be beaming towards us and the impact
parameter $\beta$ is chosen from a simple flat distribution
in the range $-\rho \leq \beta \leq \rho$. Finally, we assume
that the magnetic inclination angle $\alpha$ is randomly distributed, that is
\begin{equation}
  \alpha = \arccos q,
\end{equation}
where $q$ is a random number drawn from a flat distribution
in the range $0<q<1$. With these quantities defined, the observed
pulse width $W$ is found using the following geometrical
relationship (see, e.g., Gil et al.~1984):
\begin{equation}
\sin^2 \left( \frac{W}{4} \right) = 
\frac{ \sin^2 \left( \frac{\rho}{2} \right) - 
       \sin^2 \left( \frac{\beta}{2} \right) }{
       \sin\alpha \cdot \sin ( \alpha + \beta ) }.
\end{equation}

To compute the expected DM and scatter broadening effects on each
pulse, we use the NE2001 electron density model~\citep{NE2001}. Note
that, since we are primarily concerned with the distant population of
highly dispersed pulsars in these simulations, we do not attempt to
account for interstellar scintillation. Finally, to allow us to
extrapolate the 1400-MHz luminosities to other survey frequencies in
the next step, we make the reasonable assumption that pulsar spectra
can be approximated as a power law \citep{Lorimer95} and assign each
pulsar a spectral index drawn from a normal distribution with mean of
--1.6 and standard deviation 0.35.

As described by~\citet{Lorimer2006}, when generating the normal pulsar
population according to the above criteria, we keep track of the
number of pulsars detectable by the detailed model of the Parkes
Multibeam Pulsar Surveys of the Galactic plane~\citep{mlc+01} and at
high latitudes~\citep{bjd+06}.  Our simulations terminate when the
numbers of model detectable normal pulsars matches the 1005 pulsars
found by these surveys. This results in an underlying sample of
120,000 model pulsars.

To model the MSP population, we follow the same approach and
assumptions as for the normal pulsars described above, but with two
exceptions: (i) model $z$ heights are chosen from an exponential
distribution with a mean of 500~pc~\citep{Cordes97}; (ii) the
underlying period distribution used is taken from a recent study of
the MSP population (Lorimer et al.~2009; in preparation). Note that
implicit in this approach is the notion that the luminosity functions
for normal pulsars and MSPs are identical; this is consistent with
earlier results~\cite{lml+98}.

In addition to the Parkes Multibeam Surveys considered above for the
normal pulsars, we also make use of the Parkes 70-cm pulsar survey of
the southern sky \citep{mld+96,lml+98} and two intermediate latitude
surveys carried out by the Parkes multibeam system
\citep{ebsb01,jbo+07}.  Our simulations are normalized such that they
terminate when the total of 49 MSPs detected by these surveys is
reached. This results in an underlying sample of 23\,000 model MSPs.

Once the pulsar population was determined, we performed simulations to
find the number of pulsars FAST would detect as a function of
frequency. For this purpose, we allowed the centre frequency to range
from 400\,MHz to 1.4\,GHz. The bandwidth was kept at one third of the
centre frequency, the observation time per pointing was kept constant
at 600\,s and the gain was kept constant at 16.5\,KJy$^{-1}$. The
  frequency-dependent sky temperature was added to the system
  temperature. Considering the beamwidth of about 3.4 arcminutes at
1.4\,GHz, we limited the surveys to Galactic latitudes of either
between $\pm 5^\circ$ or between $\pm 10^\circ$. Further, we simulated
the number of pulsars FAST will detect as a function of observation
time. All detected pulsars include already known pulsars.

\subsection{Simulation Results}
\begin{figure}[htb]
\centering
\includegraphics[width=0.35\textwidth, angle=-90]{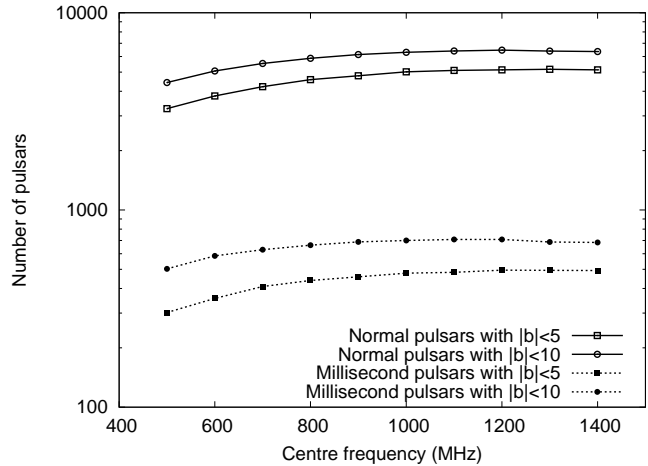}
\caption{Number of normal pulsars and MSPs detected in
  simulations of FAST surveys of the Galactic plane as a function of
  observation frequency. The surveys were limited in Galactic
  longitude by $20^\circ<\mathrm{l}<90^\circ$. The solid lines and the
  dotted lines show the number of normal pulsars and millisecond
  pulsars, respectively. The squares indicate a survey region
  limited in Galactic latitude by $|\mathrm{b}|<5^\circ$. The circles
  indicate a survey region limited in Galactic latitude by
  $|\mathrm{b}|<10^\circ$. The observation time was kept constant at
  600\,s and the bandwidth was kept at one third of the
  observation frequency. 
  }
\label{fig:Frequency}
\end{figure}

\begin{figure}[htb]
\centering
\includegraphics[width=0.35\textwidth, angle=-90]{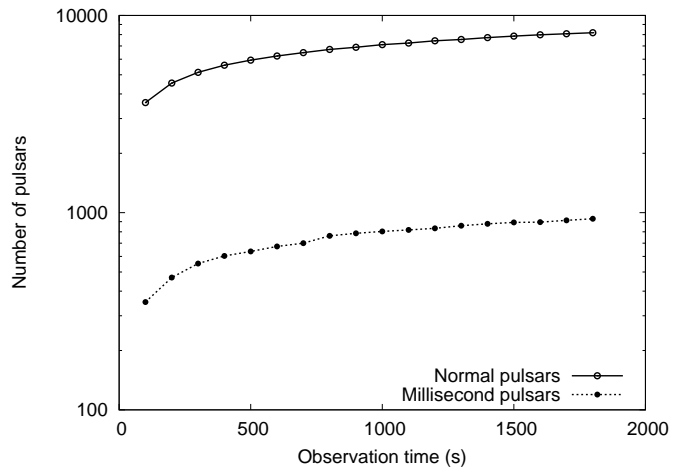}
\caption{Number of normal pulsars and MSPs detected in
  simulations of FAST surveys as a function of observation time per
  pointing. The surveys were limited in Galactic latitude and
  longitude by $|\mathrm{b}|<10^\circ$ and
  $20^\circ<\mathrm{l}<90^\circ$. The solid line and the dotted line
  show the number of normal pulsars and MSPs,
  respectively. The centre frequency is 1.315\,GHz with a bandwidth of
  400\,MHz.}
\label{fig:ObservationTime}
\end{figure}

Fig.~\ref{fig:Frequency} shows the number of normal pulsars and MSPs
that were detected in the simulation of the Galactic plane as a
function of centre frequency. The figure shows that for both normal
pulsars and MSPs, the number of detected pulsars is reasonably stable
from centre frequencies of 1 to 1.4\,GHz. Similar results are
found for observation times of 120 and 1\,800 seconds.
The number of detected pulsars as a function of observation time is
shown in Fig.~\ref{fig:ObservationTime}.  This figure shows that,
while the number of detected pulsars continues to climb with
increasing observation time, the rate of increase slows significantly
after times of about 600 s.

The 19 beams in the multibeam receiver of FAST have a total FoV of
about 0.061\,deg$^2$. A possible future PAF with over 100 beams will
have a total FoV of 0.32\,deg$^2$. Table~\ref{tab:speed} shows the
total survey time for different observation times per pointing, for
different survey regions for both a 19-beam and a 100-beam
receiver. The total survey time is given in days assuming eight hours
of observation time per day.
\begin{table*}
  \caption{Total survey time for different FAST surveys. Except for
    model 7, the survey region is limited in Galactic longitude by
    $20^\circ<\rm{l}<90^\circ$. The survey time is expressed in number
    of days with eight hours of observation. The number of detected
    normal and millisecond pulsars are rounded to the nearest 100, and
    10, respectively. The number of known pulsars has been obtained
    from the ATNF Pulsar Catalogue~\citep{Manchester05}.}
  \begin{tabular}{lrrrrrrr}
\hline
\hline
Parameter             & Model 1 & Model 2 & Model 3 & Model 4 & Model 5 & Model 6 & Model 7 \\
\hline
Survey region         & $|\rm{b}|$$<$$5^\circ$   & $|\rm{b}|$$<$$10^\circ$ & $|\rm{b}|$$<$$5^\circ$ & $|\rm{b}|$$<$$10^\circ$   
                      & $|\rm{b}|$$<$$5^\circ$   & $|\rm{b}|$$<$$10^\circ$  & Visible Sky \\
Pointing time (s)                        & 600    & 600    & 600    & 600    & 1\,800 & 1\,800 & 120 \\
Number of beams                          &  19    & 19     & 100    & 100    & 100    & 100    & 100 \\
Number of normal pulsars detected        & 5\,100 & 6\,300 & 4\,200 & 5\,200 & 5\,700 & 7\,000 & 4\,500 \\ 
Number of MSPs detected                  & 470    & 670    & 380    & 550    & 550    & 770    & 500  \\ 
Currently known normal pulsars in region & 352    & 418    & 352    & 418    & 352    & 418    &  662 \\  
Currently known MSPs in region           & 14     & 20     & 14     & 20     & 14     & 20     &  62  \\  
Total survey time (eight hour days)          &  229    &  454    &  44     & 86     &  131    &  259    &  296 \\
\hline
  \end{tabular}
  \label{tab:speed}
\end{table*}
With the initial 19-beam receiver about 5\,200 previously unknown pulsars, including
460 MSPs, can be found by searching the region with
$20^\circ<\mathrm{l}<90^\circ$ and $|\mathrm{b}|<5^\circ$, using
10-minute pointings. This survey would take just over 200 days
to complete.  Using the 100-beam receiver to survey the same region
with 30-minute pointings, 5\,900 previously unknown pulsars can be found in
130 days, including about 540 MSPs. Doubling the survey
area to $|\mathrm{b}|<10^\circ$ will add about 1\,100 pulsars using
10-minute pointings and about 1\,450 pulsars using 30-minute
pointing. The 100-beam receiver will also enable an all-sky
survey. Using 2-minute pointings, 4\,300 previously unknown pulsars can be
found in 300 days, including about 440 MSP's.

\begin{figure*}[htb]
\centering
\includegraphics[width=0.34\textwidth, angle=-90]{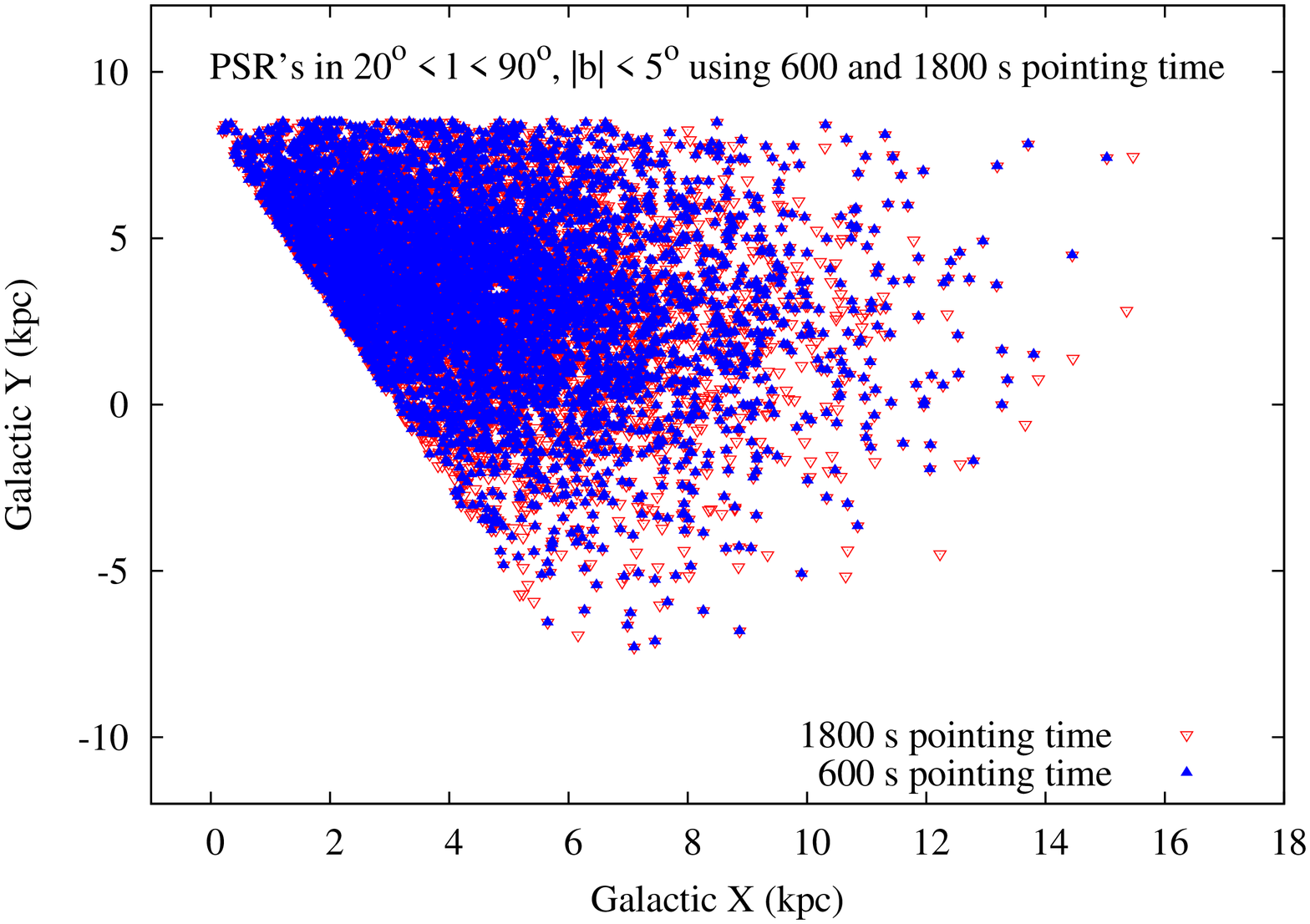}
\includegraphics[width=0.34\textwidth, angle=-90]{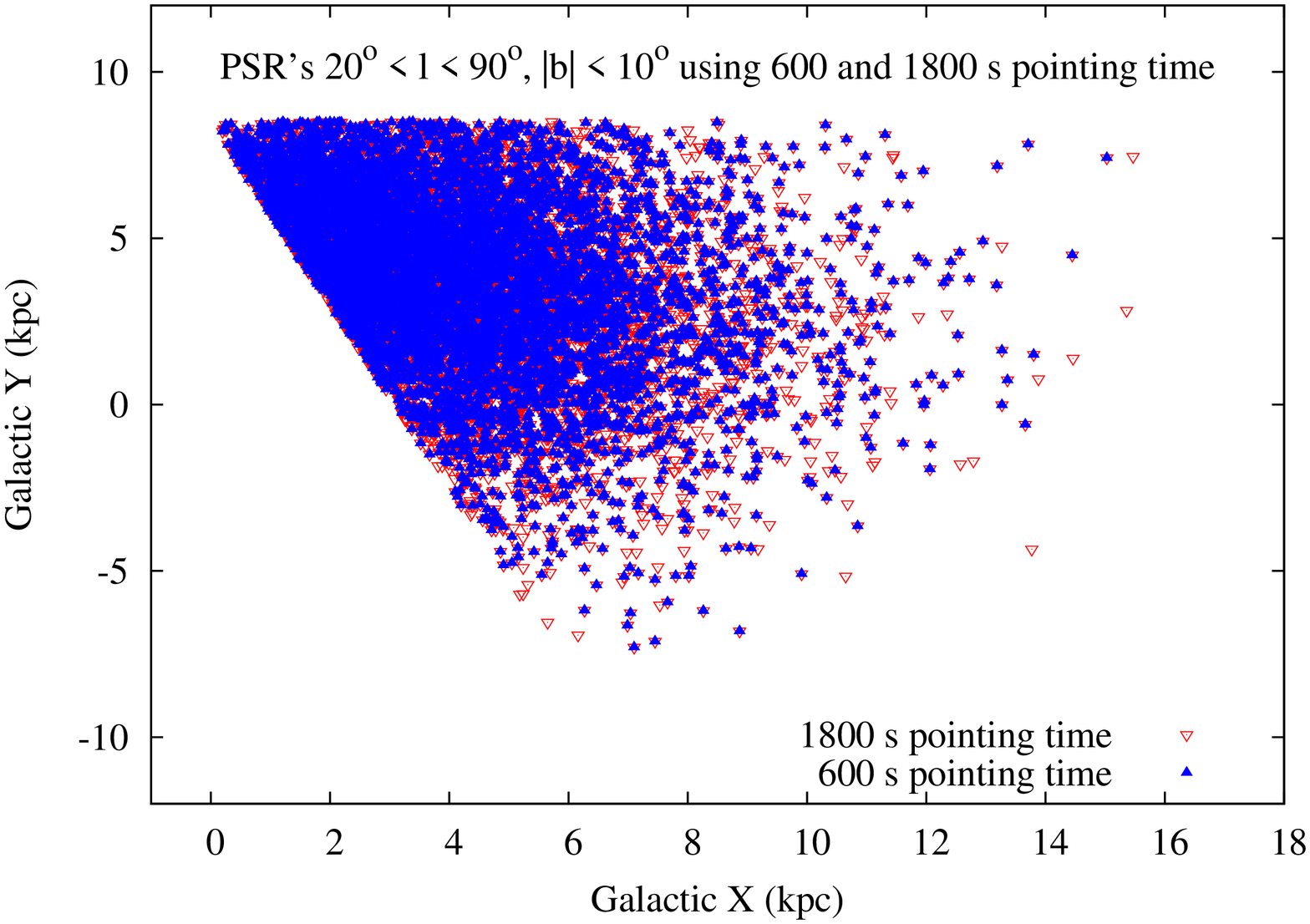}\\
\includegraphics[width=0.34\textwidth, angle=-90]{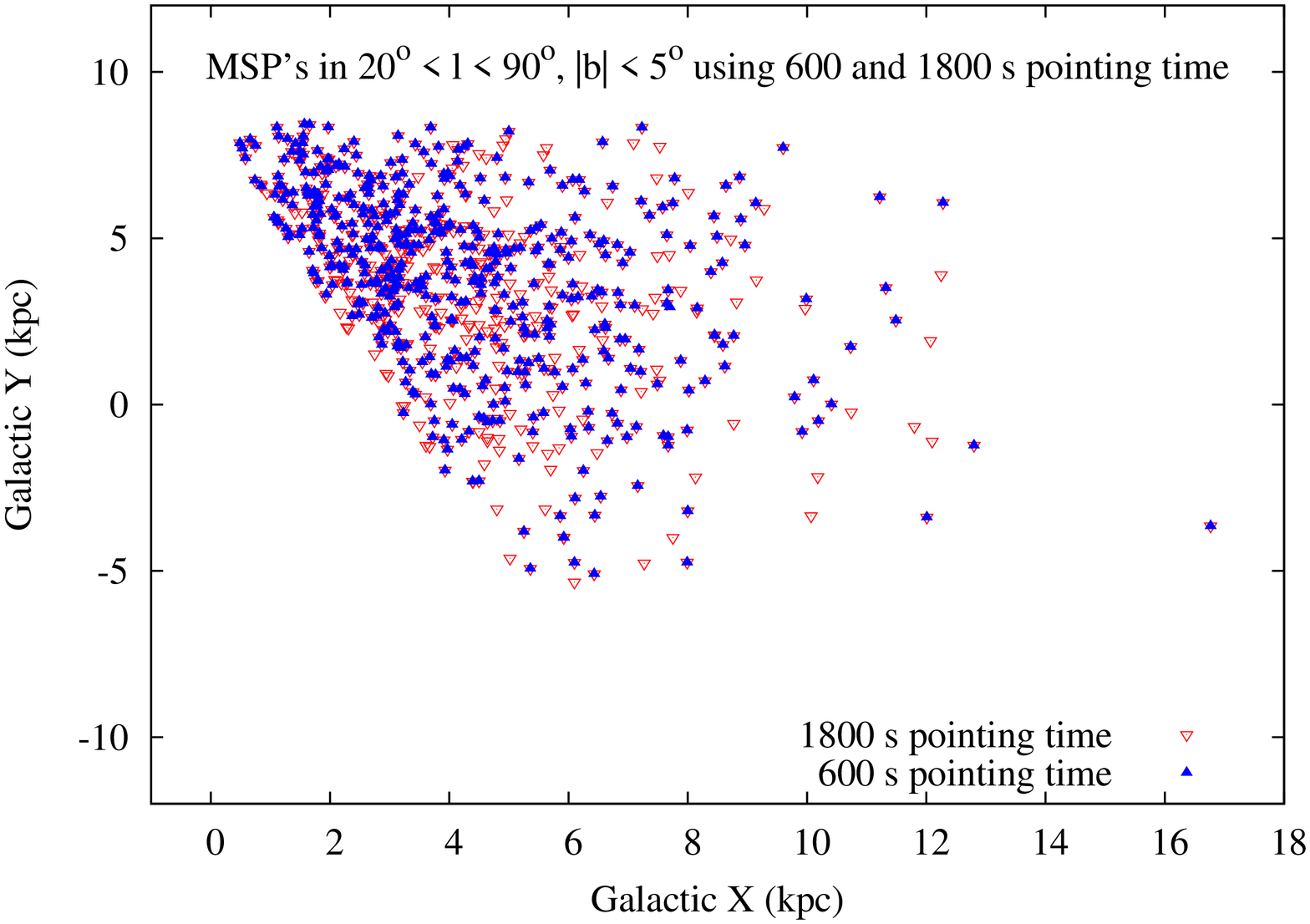}
\includegraphics[width=0.34\textwidth, angle=-90]{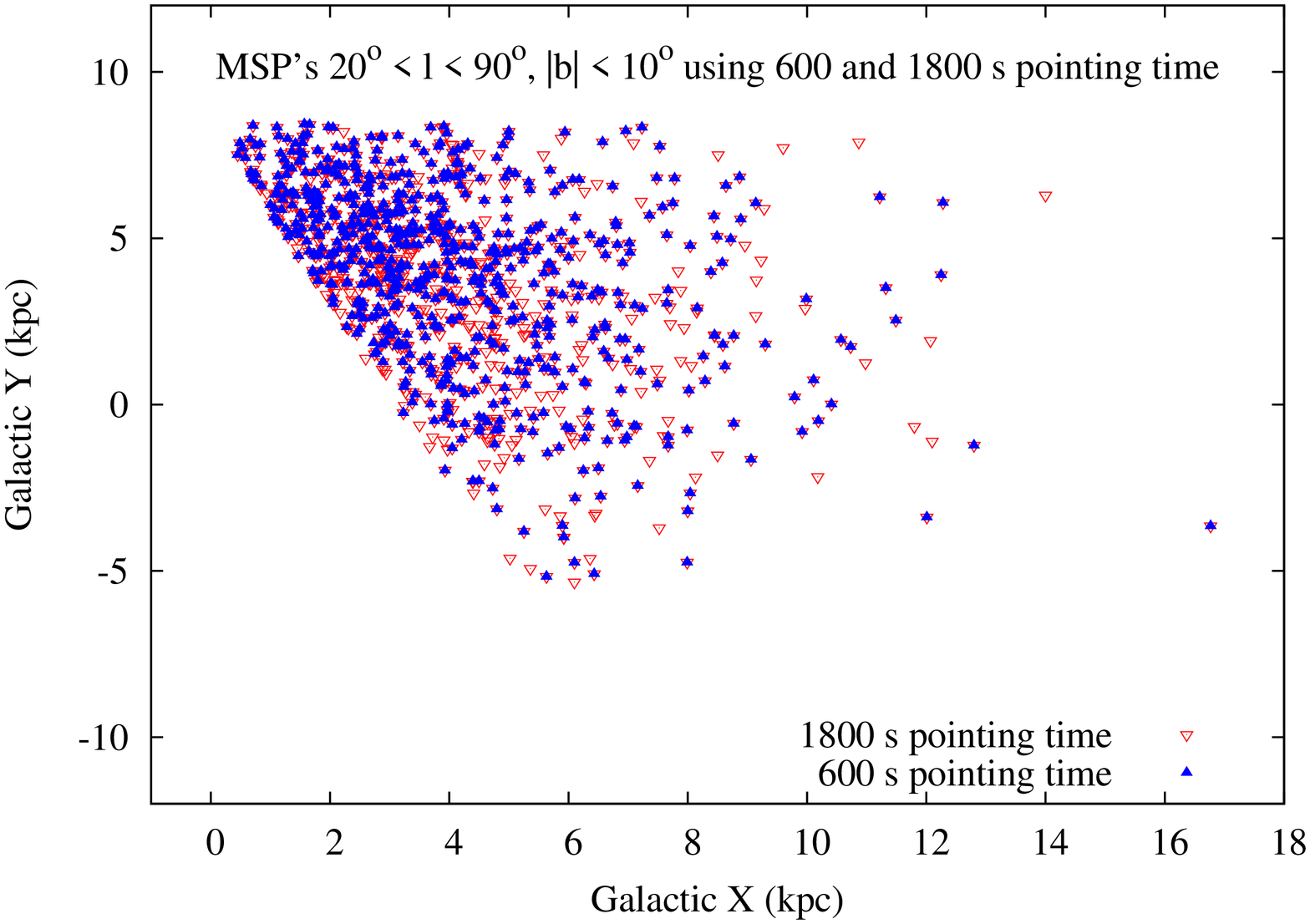}\\
 \includegraphics[width=0.34\textwidth, angle=-90]{eps/11939f07.eps}
\caption{Distribution of detected normal pulsars and MSPs for
  different models. The top left plot shows the distribution of
    normal pulsars from models 1 (similar to model 3) and 5. The top
    right plot shows the distribution of normal pulsars from models 2
    (similar to model 4) and 6. The two plots in the the middle show
    the distribution of millisecond pulsars from the same models as
    the plot directly above. The bottom plot shows a Hammer-Aitoff
    projection of the distribution of normal pulsars and millisecond
    pulsars from model 7 in galactic coordinates. The black and white
    striped boxes mark the areas defined by
    $20^\circ<\rm{l}<90^\circ$, $|\rm{b}|$$<$$5^\circ$ and
    $|\rm{b}|$$<$$10^\circ$.}
\label{fig:Scatter}
\end{figure*}

Fig.~\ref{fig:Scatter} shows the distribution of detected pulsars for
all the models. The top two plots show the normal pulsars that
  are detected in the models that are limited to the Galactic
  plane. The plots directly below show the millisecond pulsars that
  are detected in the same models. The bottom plot shows the all-sky
survey.

\subsection{Survey of M31 and M33}
The great sensitivity of FAST will permit searches for pulsars in
other galaxies. We can estimate the number of pulsars that FAST can
detect in a survey of M31 and M33, both visible for about six hours
per day from FAST. Table~\ref{tab:M31M33} shows the parameters of
these galaxies.
 \begin{table*}[htb]
  \caption{Parameters of M31 and M33.}
  \begin{tabular}{lllll}
    \hline
    \hline
    Galaxy & Declination           & Distance (kpc)               & Mass (M$_\odot$)                             & Angular extent \\
    \hline
    M31    & 41$^\circ$ 16$\arcmin$ & 772$\pm$44 \citep{Ribas05}   & $8.2\times 10^{11}$ \citep{Seigar08}         & 3\,deg$^2$\\
    M33    & 30$^\circ$ 39$\arcmin$ & 964$\pm$54 \citep{Bonanos06} & $\gtrsim 5\times 10^{10}$ \citep{Corbelli00} & 0.8\,deg$^2$\\
    \hline
  \end{tabular}
  \label{tab:M31M33}
\end{table*}
%
%
The minimum detectable flux density of a pulsar in a survey is
  given by:
\begin{equation}
S_{\rm min}=\frac{{\rm S/N}\,(T_{\rm sys}+T_{\rm sky})}{G\sqrt{n_{\rm p}t_{\rm obs}B}}\sqrt\frac{W}{P-W} {\rm mJy},
\end{equation}
where S/N is the signal-to-noise ratio required for a detection,
$T_{\rm sys}$ is the system noise temperature (K),  $T_{\rm sky}$
  is the sky temperature (K), $G$ is the gain of the telescope
(K\,Jy$^{-1}$), $n_{\rm p}$ is the number of polarisations, $t_{\rm
  obs}$ is the observation length (s) per pointing, $B$ is the
observation bandwidth (MHz), $P$ is the period of the pulsar (s) and
$W$ is the equivalent width (s) \citep{lk05}. For an observation
frequency of 1.315\,GHz and S/N = 9, $T_{\rm sys}$ = 20\,K,
  $T_{\rm sky}$ = 1.4\,K for M31 and $T_{\rm sky}$ = 1.1\,K for M33,
$G$~=~16.5\,K\,Jy$^{-1}$, $n_{\rm p}$ = 2, $t_{\rm obs}$ = 6 hours,
$B$ = 400\,MHz and assuming an equivalent width of 5\% of the period
we obtain a minimum detectable flux density of 600\,nJy. This leads to
a minimum detectable luminosity of 360 and 560\,mJy\,kpc$^2$ for a
pulsar in M31 and M33, respectively. There are at least 55 pulsars
known in our Galaxy with a higher luminosity at 1.4\,GHz than
360\,mJy\,kpc$^2$ and 32 pulsars higher than
560\,mJy\,kpc$^2$. If we correct for the mass ratio between our Galaxy
and the galaxies M31 and M33, then we can expect of order 80 pulsars
to be detected by a FAST survey of M31 and about 3 pulsars in
M33. However, the supernova rate per unit mass in our Galaxy is
thought to be higher than in M31 and considerably lower than in
M33~\citep{Berkhuijsen84, Gordon98}. Also, since the angular extent of
M33 is about 1 quarter that of M31, the number of pulsars found per
pointing might not differ much between M31 and M33. We estimate that
the total number of pulsars found in a FAST survey of both M31 and M33
would lie between 50 and 100. Finding such a number of pulsars would
allow studies of the pulsar population in these galaxies and would
provide a valuable probe of the inter-galactic medium. The combined
angular extent of M31 and M33 is 3.8\,deg$^2$, which means about
1\,500 beams are required. With a 19-beam receiver, this would require
79 pointings each of duration six hours, leading to a total of
470~hours of observing time.

\section{Data processing}
 
The total amount of data from a FAST pulsar survey is given by
\begin{equation}
{\cal D}_{total} = Bt_{obs}\frac{\Omega_{survey}}{\rm{FoV_{beam}}}N_{pol}\frac{N_{bits}}{8}\rm{\,Bytes},
\label{eq:Data}
\end{equation}
where $B$ is the bandwidth, $t_{obs}$ is the observation time per
pointing, FoV$_{\rm beam}$ is the FoV of 1 beam, $\Omega_{\rm survey}$ is the
total field of the survey, $N_{\rm pol}$ is the number of polarisations
and $N_{\rm bits}$ is the number of bits used in the digitisation.
\begin{figure}[htb]
\centering
\includegraphics[width=0.35\textwidth, angle=-90]{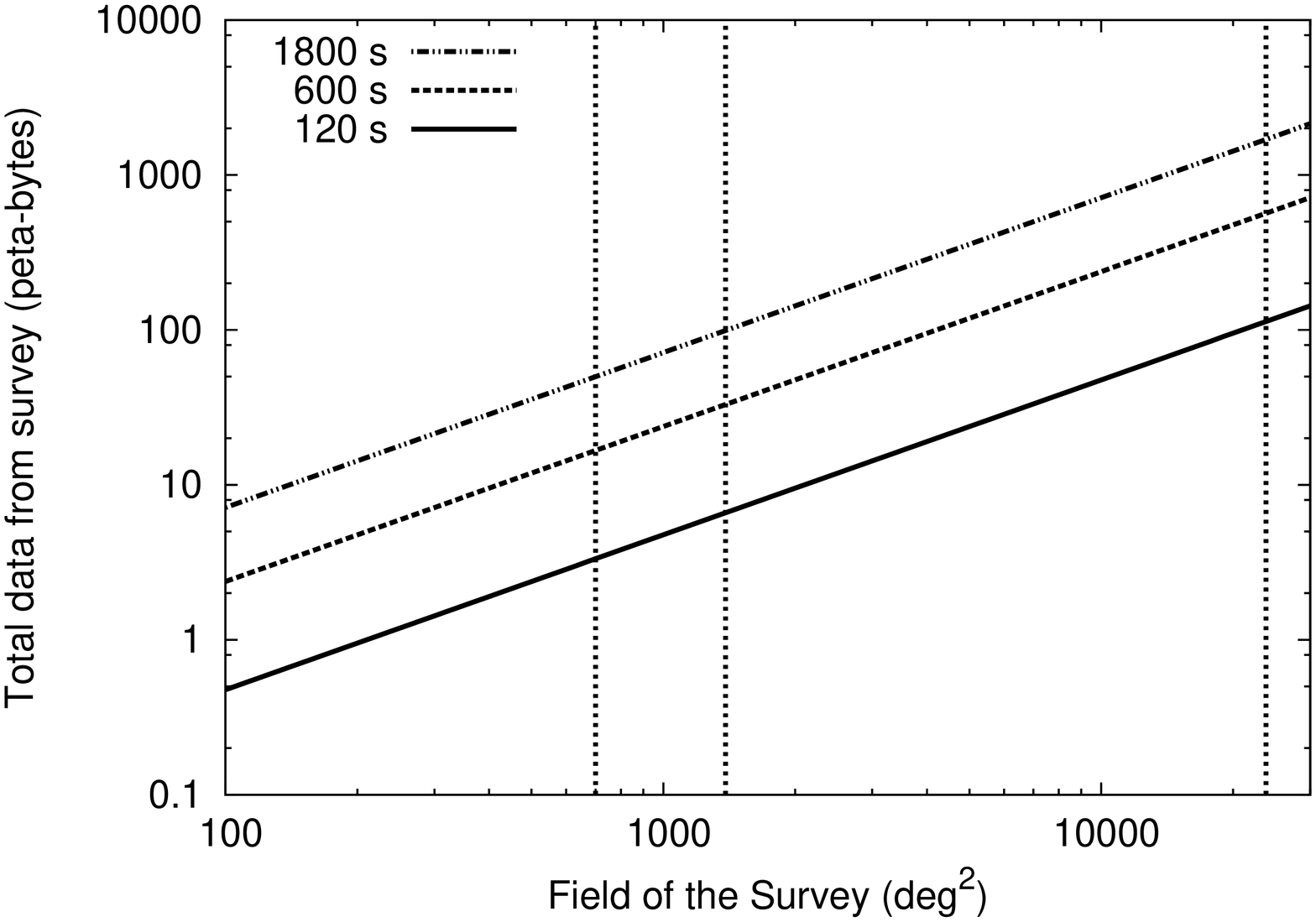}
\caption{Total amount of data from a FAST pulsar survey as a function
  of the total survey field. The calculation assumes a bandwidth of
  400\,MHz, a FoV$_{beam}$ of 0.0025 deg$^2$, 1 polarisation and 2 bits
  per sample. The three different lines correspond to observation
  times of 120, 600 and 1\,800 seconds. The dotted vertical lines
  indicate from left to right the survey fields given by
  $20^\circ<\rm{l}<90^\circ$ and $|\rm{b}|$$<$$5^\circ$,
  $20^\circ<\rm{l}<90^\circ$ and $|\rm{b}|$$<$$10^\circ$ and the
  entire visible sky.}
\label{fig:Data}
\end{figure}
Figure~\ref{fig:Data} shows the total amount of data from a FAST
pulsar survey as a function of the total survey field for different
observation times. We have assumed a bandwidth of 400\,MHz, a
FoV$_{beam}$ of 0.0025 deg$^2$, 1 polarisation and 2 bits per sample.
Table~\ref{tab:data} shows the total amount of data for the seven
different survey models from table~\ref{tab:speed}.
\begin{table*}
  \caption{Total amount of data from different survey models (see
    table~\ref{tab:speed}). Except for model 7, the surveys are limited
    in Galactic longitude by $20^\circ<\rm{l}<90^\circ$.}
  \begin{tabular}{lrrrrrrr}
\hline
\hline
Parameter             & Model 1 & Model 2 & Model 3 & Model 4 & Model 5 & Model 6 & Model 7 \\
\hline
Survey region         & $|\rm{b}|$$<$$5^\circ$   & $|\rm{b}|$$<$$10^\circ$ & $|\rm{b}|$$<$$5^\circ$ & $|\rm{b}|$$<$$10^\circ$   
                      & $|\rm{b}|$$<$$5^\circ$   & $|\rm{b}|$$<$$10^\circ$  & Visible Sky \\
Pointing time (s)     & 600    & 600    & 600    & 600    & 1\,800 & 1\,800 & 120 \\
Total amount of data (peta-bytes) & 16.7 & 33.0 & 16.7 & 33.0 & 50.0 & 99.0 & 113 \\
\hline
  \end{tabular}
  \label{tab:data}
\end{table*}

The data acquisition rate from a pulsar survey sampling the detected
outputs of a filterbank of total bandwidth $B$ and channel bandwidth
$\Delta\nu$ is given by:
\begin{equation}
  {\cal D}_{rate} = N_{\rm beams}\frac{1}{t_{\rm samp}}\frac{B}{\Delta\nu}\frac{N_{\rm bits}}{8}\mathrm{\,Bps},
  \label{eq:DR}
\end{equation}
where $N_{\rm beams}$ is the number of beams in the receiver, $t_{\rm
  samp}$ is the sampling time and $N_{\rm bits}$ is the number of bits
used in the digitisation. $\Delta\nu$ can be estimated by demanding
that the dispersion smearing within the frequency channel does not
exceed the effective time resolution, given by the sampling time and
the scattering time:
\begin{equation}
  \Delta\nu(\mathrm{GHz}) \le \frac{\sqrt{t_{\rm
        samp}^2(\mu\mathrm{s})+t_{\rm scatt}^2(\mu\rm{s})}\nu_{\rm
      min}^3(\mathrm{GHz})} {8.3\times10^3{\rm DM}},
\label{eq:deltanu}
\end{equation}
where $\nu_{\rm min}$ is the minimum (lowest) frequency in the
observation frequency band, DM is the dispersion measure and $t_{\rm
  scatt}$ is the scattering time. Following \citet{Bhat04}, we relate
$t_{\rm scatt}$ to the DM and observing frequency as follows:
\setlength\arraycolsep{2pt}
\begin{eqnarray}
\lefteqn{ \log_{10}(t_{\rm scatt}(\rm{ms})) = -6.46 + 0.154\log_{10}({\rm DM}) + \phantom{1.07(\log_{10}{\rm DM})} }\nonumber \\ 
                                &  & \qquad\qquad\qquad1.07(\log_{10}{\rm DM})^2 - 3.86 \log_{10}(\nu(\rm{GHz})).
\label{eq:tscatt}
\end{eqnarray}
In practice, however, the scattering time can differ from this by a
factor up to 100 \citep[see][]{Bhat04}. A smaller scattering time
leads to the requirement of smaller -- and thus more -- frequency
channels. To be on the safe side, we apply the highest frequency in
the frequency band to Eq~\ref{eq:tscatt} and divide the resulting
scattering time by 100. Substituting $t_{\rm scatt}$ from
Eq.~\ref{eq:tscatt} in Eq.~\ref{eq:deltanu} and taking $\nu_{\rm min}$
= 1.115\,GHz, leads to a relationship between $\Delta\nu$ and DM with
a minimum of 42\,kHz (at DM = 428). This means that having a constant
frequency channel width of 42\,kHz over the frequency band will be
sufficient for the pulsar survey. This leads to 9\,500 frequency
channels for $B$=400\,MHz.
\begin{figure}[htb]
\centering
\includegraphics[width=0.35\textwidth, angle=-90]{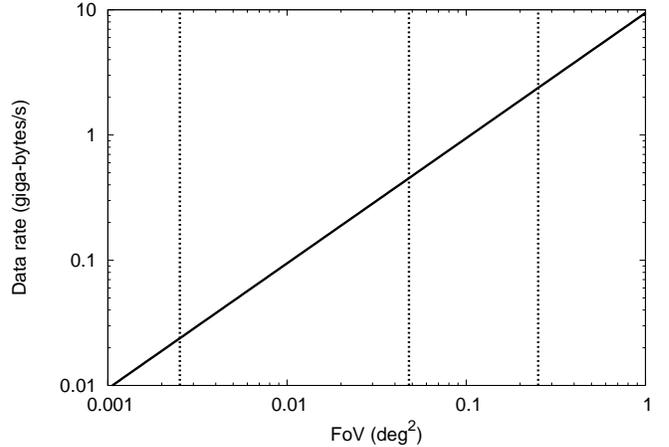}
\caption{Amount of data per second from a FAST pulsar survey as a
  function of FoV. The calculation assumes a sampling time of
  100$\mu$s, a frequency range of 1.115 to 1.515\,GHz and 2 bits per
  sample. The dotted vertical lines indicate from left to right the
  FoV of 1, 19 and 100 beams, respectively.}
\label{fig:Datarate}
\end{figure}
Figure~\ref{fig:Datarate} shows the data rate from a FAST pulsar
survey as a function of FoV. We have assumed 100$\mu$s sampling time,
a frequency range of 1.115 to 1.515\,GHz and 2 bits per sample. For
the initial 19-beam receiver, the data rate is just under 0.5 GB per
second. For 100 beams it is about 2.4 GB per second.

\begin{figure}[htb]
\centering
\includegraphics[width=0.35\textwidth, angle=-90]{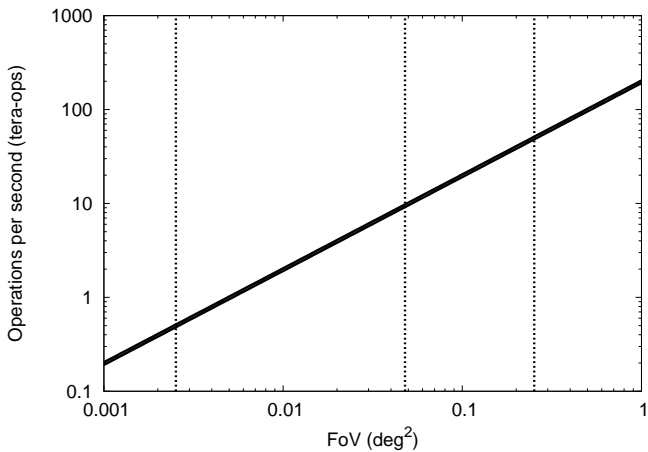}
\caption{Required computation power for real-time searching for
  possibly accelerated fast and slow pulsars in the data from a FAST
  pulsar survey as a function of FoV. The calculation assumes a
  sampling time of 100$\mu$s, a frequency range of 1.115 to
  1.515\,GHz, 2 bits per sample and a DM$_{\rm max}$ of 1\,000. The
  dotted vertical lines indicate from left to right the FoV of 1, 19
  and 100 beams, respectively.}
\label{fig:Analysis}
\end{figure}

The number of operations required to search these data for normal
pulsars, millisecond pulsars and accelerated pulsars in binary systems
is approximately
\begin{equation}
 N_{\rm oa}=N_{\rm DM}N_{\rm acc}\times5N_{\rm samp}\log_2(N_{\rm samp}),
\end{equation}
where $N_{\rm DM}$ is the number of DM-trial values and $N_{\rm
  acc}$ is the number of trial accelerations. Fig.~\ref{fig:Analysis}
shows the number of operations per second required to perform a real
time analysis of a FAST pulsar survey, assuming 100 trial
accelerations, a sampling time of 100\,$\mu$s, an observation time of
600\,s and DM$_{\rm max}$ = 1000\,cm$^{-3}$\,pc. $N_{\rm DM}$ was
obtained numerically by demanding that the total time delay within the
frequency band between two consecutive DM-trial values is equal
to twice the effective time resolution, given by the sampling
time and the scattering time of Eq.~\ref{eq:tscatt}. This leads to
$N_{\rm DM}$ = 4\,280.  For the initial 19-beam receiver, the
required computation power for real time analysis is 9 tera-ops. For
100 beams it becomes 48 tera-ops. Contributions from de-dispersion,
harmonic folding and possibly other processes would increase these
values by a factor less than two. However, it is not unreasonable
  to store all the data from a FAST pulsar survey and to perform the
  analysis off-line. This would reduce the required computation power
  significantly.

\section{Timing}

In addition to being a very powerful search machine, FAST will also
excel in the high-precision timing of pulsars. For a given pulsar, the
timing precision roughly scales with the S/N of the observed pulse
profile \citep{lk05}, and hence with telescope sensitivity. Until the
Square Kilometre Array comes on-line (\S 5)
FAST will have the largest sensitivity of all
existing radio telescopes and will also be able to see a larger
fraction of the sky than the Arecibo telescope. Consequently, FAST
will be the telescope of choice for a number of high-precision timing
experiments. While the Northern location of FAST means that it is
unlikely to contribute significantly to the timing of, say, the Double
Pulsar (PSR~J0737$-$3039A/B, Burgay et al.~2003, Lyne et al.~2004),
\nocite{bdp+03,lbk+04} a major contribution of FAST will be the
participation in the global Pulsar Timing Array efforts.

High-precision timing of a network of pulsars allows detection of a
gravitational wave background by correlating the timing residuals of
the pulsars \citep[e.g.][]{Jenet06}. Current efforts include PPTA
(Parkes Pulsar Timing Array), EPTA (European Pulsar Timing Array) and
NANOGrav (North American Nanohertz Observatory for Gravitational
Waves).  In terms of combination of sky coverage and sensitivity, FAST
will only be matched by the European LEAP efforts (Large European
Array for Pulsars), where the European 100-m class telescopes are
combined to obtain Arecibo-like sensitivity for pulsar timing.
  This effort, however, has to overcome the challenge of coherently
  combining the signals between telescopes up to 2\,000 kilometres
  apart.

\subsection{Pulsar Timing Array experiment}

Following earlier work by~\citet{det79} and \citet{Kaspi94}, pulsar
timing is now routinely used to constrain the amplitude of a
gravitational wave (GW) background.  Currently the most stringent
limits have been placed by \citet{Jenet06} from observations of seven
MSPs with the Parkes 64-m radio telescope \citep[cf.][]{Manchester08}
which constrain $\Omega_{\rm GW}$, the ratio of the energy density of
the GW background to the closure density to be $\Omega_{\rm GW}h^2 <
1.9\times 10^{-8}$, where $h$ is the dimensionless Hubble constant
defined as $H_0\equiv h\times 100$\,km\,s$^{-1}$\,Mpc$^{-1}$. The
sensitivity that is needed to achieve a GW detection depends on
details of the Galaxy merger rate, since the GW background generated
by an ensemble of super-massive black holes distributed throughout the
Universe is expected to be the strongest signal to be detected. The
signal originating from such sources is expected at a level of
$\Omega_{\rm GW}h^2 \sim 2\times 10^{-10}$ \citep{Jaffe03, Wyithe03,
  Enoki04}.  Simulations suggest that the timing of 20 pulsars with a
timing precision of 100\,ns over five years would lead to a
sensitivity of $\Omega_{\rm GW}h^2 < 6.6\times 10^{-11}$ and hence to
a first direct detection of a GW background \citep{Jenet06}. While
this is difficult with current technology, FAST will have the
sensitivity and sky coverage to make a major contribution to achieving
this goal.

Here we estimate the amount of time needed to time the brightest MSPs,
found by FAST, to high precision.  For this simulation we choose model
2 from Table~\ref{tab:speed}, which provides us with 670 MSPs. We take
the 50 brightest of these and calculate the amount of time
required for a S/N of 500, taking into account a conservative
profile stabilisation time of five minutes.  This leads to 24
hours to time all 50 pulsars once. These numbers do not include
accompanying low-frequency observations to allow correction for
variations in interstellar dispersion. As expected, we found that the
FoV provided by 19 beams had no impact on the timing performance, as
there was no instance where two or more pulsars could be timed
simultaneously.

\section{FAST versus the SKA}

The Square Kilometre Array (SKA) is a planned multi-purpose radio
telescope with a collecting area approaching 1 million square
metres. Although the exact design is not yet determined, it is likely
to consist of aperture arrays to cover frequencies from 70\,MHz up to
500 or even 800\,MHz and 15-metre dishes for the higher
frequencies. About half of the elements will be placed within a
5-kilometre core of the SKA, the rest will be placed on spiral arms
extending several thousands of kilometres outwards. \citet{Smits09}
have studied the performance of the SKA concerning pulsar surveys and
timing. Here we will discuss the differences between these findings
and those of the current paper.

A pulsar survey with FAST is straightforward compared to an SKA pulsar
survey. With the initial receiver, no beam-forming is required. The
future PAF does require beam-forming, but it does not need the vast
computational power of several peta-ops required for beam-forming the
SKA. Another benefit for FAST is that the data do not need to be
transported over long distances; the data processing can be performed
close to the dish. Also, the data rates and required computation power
for a real time acceleration search are much lower for the FAST
survey. This is mostly due to the smaller FoV of FAST, which is 10 to
50 times less than the FoV for an SKA pulsar survey, but also because
these data rates and computational power scale with the square of the
telescope diameter. The diameter of FAST is given by the illuminated
aperture of 300 metres. For the SKA it is determined by the largest
baselines between the elements used in the survey.  \citet{Smits09}
estimate that the limits in computational power will restrict these
baselines to 1 kilometre. Because of this limitation, the telescope
sensitivity of a pulsar survey with the SKA is about
2\,000\,m$^2$K$^{-1}$ which is equal to that of FAST. However, the
larger FoV of the SKA will not only enable a survey of a much larger
portion of the sky, but also allow for a longer dwell time per
pointing. The actual FoV of the SKA depends on the final design, but
at the very least it will be equal to the FoV of the 15-metre dishes
which at 1.4\,GHz is about 0.64\,deg$^2$.  For comparison, a
reasonable assumption would be that an SKA survey has a dwell time of
40 minutes versus 10 minutes for a FAST survey. This makes the survey
sensitivity of the SKA twice that of FAST. Compared to the initial
19-beam receiver of FAST, a pulsar survey with the SKA would be about
2.5 times faster, but with the 100-beam PAF, the survey speed of FAST
would be twice that of the SKA. Moreover, because the elements of the
SKA are fully steerable, the SKA will be able to survey a much larger
part of the sky and with its latitude of about --30$^\circ$ this
includes almost the entire Galactic plane.

Timing pulsars with the SKA can be performed with a considerable
fraction of the full collecting area, providing a sensitivity close to
10\,000\,m$^2$K$^{-1}$.  Also, the large FoV of the SKA allows timing
of many pulsars simultaneously (but depending on the SKA design, this
might not be the case when using the SKA for a pulsar timing array).
Even the future 100-beam PAF of FAST only has a quarter of the FoV of
15-metre dishes and should the SKA be equipped with PAFs or
mid-frequency aperture arrays, its FoV increases significantly to
20\,deg$^2$ or even 250\,deg$^2$.

Given that FAST will be operational well ahead of the full SKA, it is
likely to be the most powerful telescope to perform pulsar science in
the next decade and will remain an outstanding instrument for pulsar
science, complementing the SKA once it is operational.

\section{Summary}

The FAST pulsar survey simulation in \S 2 shows that, despite the
small natural beam-size and limitations in zenith angle, FAST will be
a formidable instrument for finding pulsars.

A tradeoff between the number of pulsars and total observation time
can be determined from Table~\ref{tab:speed}. With the initial 19-beam
receiver about 5\,200 previously unknown pulsars, including 460
MSPs, can be found in a survey of about 230 days of the visible
part of the Galactic plane with $|{\rm b}|<5^\circ$. With a 100-beam
receiver 5\,900 previously unknown pulsars can be found in
the same region in 130 days, including about 540
previously unknown MSPs. The 100-beam receiver will also enable an
all-sky survey yielding 4\,300 previously unknown pulsars
in 300 days. Further, we estimate that a 470-hour survey of M31 and
M33 would yield between 50 and 100 extra-Galactic pulsars.

A pulsar survey with the 19-beam receiver will produce just
under 500~MB of data per second. Depending on the survey, this
  leads to a total of 17 or 33 peta-bytes. The major part of a
real-time acceleration search of these data would require 9
tera-ops. After 2014, such data rates and processing power should be
very feasible.

FAST will also be able to contribute to the existing efforts to detect
a gravitational wave background by timing a large number of MSPs to
high precision. Although the visible sky of FAST is limited to 58\% of
the entire sky, those pulsars that are visible to FAST can be timed to
great precision very quickly. A simulation in \S 4 suggests that the
50 brightest MSPs, visible to FAST can be timed to a S/N of 500 in
just 24 hours.

Compared to a pulsar survey with the SKA, FAST will match the
sensitivity of the SKA, but the large FoV of the SKA will allow for
longer dwell times and make the survey faster, at least for the
initial 19-beam configuration of FAST. Also, the SKA will be able to
survey a much larger part of the sky. However, having a single dish
rather than many elements spread over a large area significantly
reduces the required data rates and computational power. When timing
pulsars for a pulsar timing array, the SKA will have much more
sensitivity than FAST, since the SKA can use almost the full
collecting area in this case. Also, the larger FoV of the SKA will
allow timing of many pulsars simultaneously, although this will
probably not benefit the timing of the limited number of pulsars from
the timing array. Given that FAST will be operational well ahead
  of the full SKA, it will provide the best prospects for pulsar
  science in the next decade.

\acknowledgement{ We would like to thank the referee, Scott Ransom,
  for his useful suggestions and comments. The authors have made use
  of the ATNF Pulsar Catalogue which can be found at
  \url{http://www.atnf.csiro.au/research/pulsar/psrcat}.  This
  effort/activity is supported by the European Community Framework
  Programme 6, Square Kilometre Array Design Studies (SKADS), contract
  no 011938. DRL is supported by a Research Challenge Grant from West
  Virginia EPSCoR. }

\bibliographystyle{aa} 
\bibliography{11939}

\end{document}